**New techniques for imaging and identifying defects in electron microscopy**

Daniel S. Gianola[1]*, T. Ben Britton[2], Stefan Zaefferer[3]

[1] Materials Department, University of California Santa Barbara, USA;
gianola@ucsb.edu

[2] Department of Materials, Imperial College London, United Kingdom;
b.britton@imperial.ac.uk

[3] Max-Planck-Institut für Eisenforschung, Düsseldorf, Germany;
s.zaefferer@mpie.de

Defects in crystalline materials control the properties of engineered and natural materials, and their characterization focuses our strategies to optimize performance. Electron microscopy has served as the backbone of our understanding of defect structure and their interactions owing to beneficial spatial resolution and contrast mechanisms that enable direct imaging of defects. These defects reside in complex microstructures and chemical environments, demanding a combination of experimental approaches for full defect characterization. In this article, we describe recent progress and trends in methods for examining defects using scanning electron microscopy platforms, where several emerging approaches offer attractive benefits, for instance in correlative microscopy across length scales and *in situ* studies of defect dynamics.

**Introduction**

The paradigm underpinning materials science and engineering that structure controls properties and performance is determined by the governing role of defects in mediating properties ranging from mechanical strength and damage tolerance (*1–3*) to optoelectronic response (*4*) to phase transformation phenomena (*5*). Direct experimental characterization of crystalline defects dates back to the invention of electron microscopy (*6*), yet the past decade has witnessed tremendous advances in new electron imaging and diffraction-based modalities for quantifying defects and their corresponding ensembles, interactions with other microstructural features, and dynamics. In the case of extended defects such as dislocations and planar faults, pioneering developments have taken an orthogonal tack from the race for spatial resolution in electron microscopy, and instead have targeted correlative characterization techniques, temporal resolution to capture dynamics *in situ*, and statistical quantification of defect evolution and organization. This is required to deploy modern materials for emerging technologies such as additive manufacturing and optoelectronics, and those used in extreme environments, where a toolbox of materials characterization probes is necessary to advance our understanding of the links between defects and materials properties.

This article focuses on methods for identifying and quantifying defects that are amenable to scanning electron microscopy platforms, which offer versatility for multi-modal and *in situ* characterization, as well as differences in scattering physics owing to different primary electron beam energies. We highlight recent advances and trends in defect imaging and characterization using electron backscatter diffraction, electron channeling contrast imaging, and diffraction-contrast scanning transmission electron microscopy approaches. We conclude by assessing the role of the emerging interplay between multimodal microscopy and data science on our understanding of defect-property relationships in advanced materials.

**Defect characterization using electron backscatter diffraction (EBSD)**

Progressive development of EBSD has increased in sophistication and presently provides fast automated indexing of electron diffraction patterns in the SEM (for more information, see (*7*) (*8*) (*9*) and (*10*) ) For defect analysis, this has been augmented by recent developments of the HR-EBSD method initially advanced by Wilkinson and colleagues (*11, 12*) which uses direct cross correlation of EBSD patterns which enables a resolution of better than $10^{-4}$ in (relative, deviatoric) elastic strain and $10^{-4}$ rads in (relative) lattice rotation. For metallic structures, recent advances have included pattern remapping (*13, 14*) which has improved the robustness of elastic strain measurements in metals. For all these EBSD techniques, the scanning nature of the electron beam in the SEM and automated analysis of very large maps (often >>10k mapped points) at a range of length scales enables rich microstructure maps to be generated (for reviews on the EBSD technique see (*15, 16*)). The information within each measured EBSD map is rich, and the information (e.g. grain shape, orientation) can be correlated together providing insight into crystal orientation, grain boundary network. These data can be linked with other imaging modes such as backscatter and secondary electron imaging, as well as AFM (*17*) and Raman microscopy (*18*) to provide correlative approaches to understand the lattice state in materials. Through applying Nye's analysis (*19–22*), and afforded by the increased precision of the HR-EBSD approach (*23*), it is now routine to assess the storage of so called "geometrically necessary dislocations" (GNDs) which give rise to lattice curvature; furthermore, a statistical treatment of the variation in lattice shear stress distributions can be used to assess the "statistically stored dislocations" (SSDs) (*24, 25*), which are related to closely bound dislocations that do not give rise to lattice curvature e.g. dislocation dipoles and multipoles. Mapping of GND content is a popular use of the EBSD method and these quantitative maps can span a range of dislocation densities ($1\times10^{12}$ to $\sim5\times10^{15}$ dislocations per $m^2$) and length scales (from $\sim2$ nm to 10 µm step size) (*26*) which are often difficult to assess with other methods. Understanding of defects with HR-EBSD has proven popular in metals (*27–30*) , semiconductors (*31, 32*) ceramics (*33*), and geological materials (*34–36*).

The HR-EBSD and EBSD techniques provide quantitative assessment of the state of the crystal lattice and therefore they lend themselves towards direct quantitative linking with materials models, such as crystal plasticity finite element modelling. For HR-EBSD, the technique can only measure relative variations in lattice strain within each grain, which can be related to the relative stress through Hooke's law. Nevertheless, even these types of residual stresses, caused by the presence of lattice defects, are important to understand the nature of the lattice strain state during deformation (*37, 38*) or that of thin film crystal growth (*12*).

EBSD continues to evolve as a technique, most notably with ever faster detectors (modern CMOS based detectors can capture patterns at >2000 Hz) and the emergence of direct detectors (*39–41*), which provides high angular information within the patterns themselves, thus offering the potential for direct quantitative comparison with high quality dynamical simulations (*42*). This has the potential to unlock further information about the structure and nature of defects, as it may be possible to correlate the selective blurring of different diffraction bands with the nature of the dislocation structures within the interaction volume (*43*). This approach is analogous to line broadening approaches used commonly within the X-ray community (*44*) using advances in pattern analysis (*45, 46*), but note that dynamical scattering and diffraction effects and the detector physics would have to be considered.

At a smaller length scale, transmission Kikuchi diffraction (TKD) (*47*) has opened up STEM based diffraction within the SEM, where wide angle (i.e. short camera length as compared to the TEM) Kikuchi patterns are now captured routinely. These wide-angle patterns can be indexed using conventional EBSD analysis methods, and thus the microstructure of nano-crystalline materials can be now unlocked (*48, 49*). Advances in this area are likely to involve dynamic testing using *in situ* methods, including heating to observe nano-grain growth (*50*), indentation-based testing similar to the TEM (*51*), but with fast STEM based quantification of the associated diffraction patterns, and thus unlocking the identification of stress induced phase transformations and crystallographic reorientation via mechanisms such as twinning. This will take *in situ* EBSD and mechanical down a length scale further beyond micro-cantilever (*52*) and micro-pillar testing (*53–55*).

**Electron channeling contrast imaging (ECCI) of defects**

　　　Electron channeling contrast imaging (ECCI) is an SEM-based imaging technique by which extended defects, i.e. dislocations, stacking faults, grain boundaries, nano-twins and precipitates and elastic strain fields in crystalline matter can be observed directly on bulk samples through backscattered electrons (BSE). ECCI is not a new technique; EC contrast was observed and correctly interpreted in the 1960s by Coates (*56*). Its features and potential applications were treated in detail by Joy et al. (*57*). Nevertheless, the technique remained exotic because of shortcomings of the available microscopes. Only with the advent of SEMs with high-brightness thermal field emission guns, parallel illumination, sensitive backscatter detectors and versatile stages, the technique developed into a useful and competitive technique. Many research groups have since contributed to the better understanding and application of this technique, e.g. (*58*) on dislocation imaging, (*59*) on Burgers vector analysis, (*60–63*) on simulation of dislocation images and understanding of contrast formation and ourselves on the combination of EBSD with ECCI for accurate diffraction condition determination (*64*).

　　　The contrast formation principles for ECCI are very much the same as in bright field scanning transmission electron microscopy (STEM) with the important difference that one does not directly observe the diffracted electron intensity but rather the backscattered intensity which is modulated by the scattering and diffraction as the beam enters the bulk crystalline sample and subsequently escapes. As a consequence, the contrast is weaker and inverted compared to STEM. Furthermore, as the signal stems from a bulk sample (and not from a thin foil as in STEM) a blurring background signal arises from those electrons which, due to inelastic scattering processes, have left the original coherent electron wave field. The spatial resolution of ECCI is limited mainly by scan control, the width of the primary electron beam, and by its broadening inside of the sample. At optimum microscope conditions the lateral resolution is currently on the order of 8 nm (*64*). Depending on the electron imaging and sample conditions (mass and defect density) the signal originates approximately from the first 100 nm below the surface (*64*).

Figure 2(a) shows an example of an ECC image, taken from a lightly deformed high entropy alloy showing a high density of dislocations mainly arranged in planar slip bands. Some dislocations appear in conjunction with stacking faults, demonstrating a relatively low stacking fault energy; others appear in pairs likely indicating short-range ordering. Using the same principles as conventional diffraction-based TEM the crystallographic character of the defects, e.g. Burgers vector and line direction of a dislocation, can be quantified in many cases. In order to do so, however, a tool is required to accurately determine the crystal orientation and the active diffraction conditions. This can be done via EBSD (*64*, *65*), electron channeling patterns (*66*) or <span style="color:red">interpretation of contrast variation for multiple channeling contrast images (*67*).</span> The EBSD-based approach is depicted in Figure 2(b): a sample is placed in EBSD position (1) to determine the crystal orientation of a desired grain (2). The sample is then moved close to the BSE detector and is tilted to the approximate channeling conditions using the knowledge on crystal orientation; the sample tilt and rotation is then slightly adjusted to maximize contrast (3). By scanning over the sample an ECC image is obtained (4).

ECCI is advantageous as it works with bulk samples, rather than thin foils thus reducing bending and strain relaxation. Therefore, defect evolution can be observed with *in situ* or quasi *in situ* experiments with improved boundary conditions, for example during deformation, annealing or chemical modification. Furthermore it can be used on very large (up to $cm^2$) improving statistical treatments. This often overcomes issues with contrast and resolution, as compared to complementary TEM and STEM based defect analysis.

The powerful features of ECCI are illustrated in Figure 2(c-e) with an example from a study on hydrogen-dislocation interaction in a high-Mn austenitic steel sample: Figure 2(d) displays the ECC image of the dislocation field formed around a nano-indent (indenter sphere diameter 1 μm, indentation depth 100 nm). The current diffraction conditions are displayed by the active diffraction vector, ***g***. Individual dislocations are visible immediately outside of the indentation area, approximately 200 nm from the indenter center. Different slip systems are visible, reaching the surface at different distances from below the indent. The traces of the {111} slip planes, obtained through EBSD orientation measurements, are displayed in the figure as well. The fact that the indent is observed on a bulk sample makes it possible to observe many more of these indents, obtained from different grains and at different locations, as it is shown in Figure 2(c). From the dislocation fields obtained for each grain it was possible to extract, with high statistical significance, which features were systematic and which ones stochastic. In a subsequent step the sample was lightly polished to remove the indents but conserving the grains; it was then electrolytically charged with hydrogen and the same grains indented and observed again. One resulting indent is shown in Figure 2(e). Here, dislocations with extended stacking faults reach much further out than in the uncharged sample, indicating a reduction in stacking fault energy by hydrogen and an increase of dislocation density as proposed by the hydrogen enhanced local plasticity (HELP) mechanism (*68*).

**Diffraction-contrast scanning transmission electron microscopy**

Some of the earliest work on TEM included the imaging of crystalline defects (*69*), providing experimental corroboration of the theory of defects in materials. Today, TEM is often the quintessential mode of direct imaging and characterization of individual crystalline defects. Routine diffraction-contrast TEM is based on local deviations from the Bragg condition owing to defect-induced displacement fields. Conventional TEM (CTEM) approaches using parallel beam illumination allow for not only imaging of defects, but characterization of both Burgers and line vectors from dislocations as inferred from their shape in the image and from invisibility conditions in diffraction; furthermore the displacement vectors from planar faults can be characterized. Imaging of defects using higher order diffraction vectors, known as weak-beam TEM (*70*), facilitates studies of the fine structure of dislocations owing to contrast that is localized near the defect, but requires long exposure times that demand stable and drift-free conditions.

Recently, diffraction-contrast STEM has been rekindled as a promising approach to characterize crystalline defects (*71–74*), in part thanks to the advent of high-quality electron sources. The primary advantages of STEM-based defect studies in which a convergent focused probe is used, in comparison to CTEM approaches, include the suppression of auxiliary contrast features such as bending contours and thickness fringes (Figure 3(a,b)), reductions in extinction contrast from inclined dislocations, and the ability to image defects in relatively thick specimens owing to the mitigation of chromatic aberrations arising from post-specimen lenses needed in CTEM. These collective benefits offer exciting opportunities for semi-automated quantification of defect densities because of more uniform contrast (*75*). Recent studies have shown that dislocation invisibility is still applicable in STEM for Burgers vector determination, and moreover that weak-beam conditions are achievable albeit with high signal-to-noise ratios enabled by annular STEM detectors where the signal is integrated over a given acceptance angle (*72*). Taken as a whole, the advantages of STEM diffraction-contrast defect imaging additionally lend themselves very well to *in situ* experiments and tomography, making it the emerging tool of choice in modern defect-property studies.

Parallel developments in modern solid-state STEM have enabled integration with commercial SEM platforms (Figure 3(c)), augmenting their conventional functionality. In such transmission modalities, the lower primary beam energies from SEM sources (<30 kV) are appealing in applications that suffer from low contrast in weakly scattering objects and knock-on beam damage (*76*). These conditions favor studies of organic and low atomic number materials, which indeed have been the primary focus of so-called STEM-in-SEM (*77*), or transmission SEM (TSEM) (*78, 79*), techniques. Pioneering studies showed the promise of TSEM at 30 kV and below in a large-chamber SEM environment, which date several decades ago (*80–82*). Approaches reported since have emphasized the need to control the acceptance angles used for image formation since camera lengths must be set physically (such as through the use of pre-machined aperture masks (*77, 83*) and a digital micromirror device (*84*)). Modern segmented detectors available for TSEM offer a range of annuli with controlled acceptance angles (and even azimuthal segmentation) and even high-angle annular regions for mass-thickness dark field imaging. The reduction in price and improvement in quality of direct electron detectors is likely to assist further in improving TSEM.

The utility of TSEM extends well to the imaging and characterization of crystalline defects as was demonstrated in earlier work (*85–87*), with recent studies advancing this capability and understanding. Diffraction-contrast images of dislocations and stacking faults in semiconductors (*88*) and metallic materials (*78*) reproduce the distinct advantages of a convergent scanning probe and also offer practical benefits of providing statistical studies of large areas enabling high-throughput studies since many specimens can be incorporated in the large vacuum chambers. The incorporation of more sophisticated instrumentation for *in situ* studies of defect dynamics and evolution than cannot fit in the small volume of a TEM holder is a notable advantage. The multimodal detection schemes offered in a well-equipped modern SEM (such as EBSD, EDS, CL, ECCI, back-scattered detectors) pair nicely with TSEM and facilitate exciting simultaneous correlative studies. One unique aspect of TSEM for defect imaging arising from the low primary beam energies is that defect contrast is strongly localized compared with the electron energies used for TEM and STEM (~200-300 kV). As studied by Callahan *et al.*, TSEM conditions show image qualities for dislocations and stacking faults that are reminiscent of weak-beam TEM conditions, although they are formed from strong beams providing very good signal-to-noise ratios (*78*) (Figure 3(d-f)). This can be rationalized by considering that the image width of a defect is proportional to the effective extinction distance for a given diffraction vector, defined as

$$\xi_{\mathbf{g}}^{eff} = \frac{\xi_{\mathbf{g}}}{(1 + s_{\mathbf{g}}^2 \xi_{\mathbf{g}}^2)^{1/2}},$$          (1)

where $s_{\mathbf{g}}$ is the deviation parameter. Weak beam TEM narrows the defect contrast by increasing $s_{\mathbf{g}}$, whereas TSEM at 30 kV provides an approximately 3x decrease of the extinction distance $\xi_{\mathbf{g}}$ compared with 200 kV TEM, which results in narrow defect widths even in the case of strong beams (i.e. $s_{\mathbf{g}} = 0$). In addition, a smaller Ewald sphere at 30 kV (and below) implies that the deviation parameter increases more rapidly as one moves away from the Bragg condition, suggesting that the strain field near a dislocation will provide more localized contrast (Figure 3(g,h)). These collective features make defect observations using TSEM amenable to materials that possess high dislocation densities or where fine structure needs to be resolved (e.g dissociated dislocations), as well as where defect-obstacle interactions are of interest (*89*). The field of TSEM as applied to characterization of defects is nascent and offers exciting practical and fundamental benefits for materials research. Advancements in this vein are necessary, such as in navigating reciprocal space to specific diffraction conditions by making use of on-axis cameras providing diffraction patterns (*88*), as well as in advanced positioning systems analogous to those found in advanced X-ray synchrotron beamlines.

We offer the following comparison of the methods described above for defect assessment, as summarized in Table 1. Taken as a whole, both ECCI and EBSD offer the distinct advantage of being applicable to the surface of bulk materials and over large imaging areas, in comparison to CTEM analysis. EBSD provides quantitative insight owing to the collection of spatially-mapped diffraction patterns (encoding crystallographic phase, lattice strains, orientations, and rotation gradients), although drawing links to dislocation configurations and arrangements requires thoughtful inference and often recourse to models. By comparison, ECCI provides a direct means of imaging and characterizing defects, although the contrast and spatial resolution is generally poorer than the CTEM counterparts because of differences in interaction volumes. If higher spatial resolutions are needed, then transmission imaging and diffraction modalities applied to thin specimens, such as diffraction-contrast STEM (and the SEM-based version TSEM) and TKD fulfill these needs, provided the specimen preparation challenges can be overcome and any thin film effects on defect structure can be reconciled. The use of lower incident electron energies (e.g. 30 kV) used in TSEM compared with CTEM results in larger electron probe sizes, yet offers very narrow dislocation image widths with high signal-to-noise ratio, which would be beneficial in instances where dislocation densities are high or dislocation-obstacle interactions are difficult to discern. In all cases, the practical benefits of SEM environments such as ease-of-use and large chamber sizes conducive to sample throughput or *in situ* instrumentation often outweigh the fundamental limitations of the 'simpler' electron microscope.

**Table 1.** Order of magnitude comparison of defect characterization approaches.

|  | **ECCI** | **EBSD** | **TKD** | **CTEM** | **STEM** |
|---|---|---|---|---|---|
| *Lateral resolution* | 10 nm | 20…500 nm [1)] | 10 nm [1)] | 1 nm | 0.1 nm |
| *Depth of observation* | 50…100 nm [1)] | 10…30 nm [1)] | 10 nm [1)] | 100…200 nm | 20…50 nm |
| *Observable area* | $10^8$ μm² | $10^8$ μm² | $10^4$ μm² | $10^4$ μm² | $10^3$ μm² |
| *Sample type* | bulk | bulk | thin foil | thin foil | thin foil |

[1)] Depending on atomic number of sample and acceleration voltage

## Advances in algorithms and data science: toward predictive defect-property relationships

The quality of the information now obtained with state-of-the-art microscopy is high contrast and information rich. Each microscope image contains information which is a convolution of microscope conditions (e.g. beam convergence angle, scanning directions, detector contrast) and the beam-material interactions (i.e. signal modulation due how the electron interacts with the sample and the signal escapes). When diffraction patterns (and spectroscopic information) are obtained at each mapped point, the richness of the information increases further. In Figure 4, we highlight examples of further insights from correlative microscopy (multi modal, including chemistry and structure) and *in situ* microscopy (providing time) approaches. The volume and complexity of these approaches automatically lend themselves to applications of "big data" and "machine learning" approaches. In gathering data, the SEM community has been ahead of many in this regard, as automation has been key in handling and reducing the data obtained (e.g. with automated indexing of diffraction patterns) to provide easy ways to interpret micrographs and provide immediate and direct interpretation.

Deterministic analysis of SEM data is routine, i.e. where the data reduction strategy is known *a priori*, as most SEM experiments are established with a good idea of how to optimize contrast to see particular features. For instance, EBSD data reduction and analysis is made easier as new Open Source toolboxes are released that simplify and improve the common handing of typically operations, ranging from the 'trivial' tasks of plotting of mapped data and the rendering of grain boundary structures networks in 2D and 3D (as performed in MTEX (*90*) for 2D and 3D EBSD, and 3D within Dream3D (*91*)). Quantitative handling and data processing (e.g. indexing with the Hough/Radon transform) of EBSD diffraction patterns is now afforded in AstroEBSD (*92*) and this is enhanced with the establishment of a translatable description of the frames of reference used (*93*).

Forward modelling is increasingly used as greater computation power and numerical approximations makes solving of complicated electron-matter physics interactions reasonable, which makes fitting of these models tractable. For electron modalities including electron channeling and diffraction can be performed using EMSoft (*94*), which has widened the opportunity to perform high quality pattern matching based indexing of electron channeling patterns (ECPs) and EBSD. Advances in this area will likely include the use of forward models to provide, for instance, the generation of physics based templates for matching different dislocation types that thread the surface of bulk semiconductors (*95*) and improve the robustness to image noise thus realizing automated defect analysis (*96*) in industrial processes.

Statistical (often information blind) approaches are being applied as numerical tools and scripts are increasingly common. These approaches tend to provide inferred correlation but rarely causation. For defect analysis can be built using tools such as multivariant statistical analysis (MSA) and principal component analysis (PCA) to improve signal to noise and enable common microstructural features to be identified with EBSD and TKD (*97*), and future applications in the SEM community will likely build in concert with developments of HyperSpy (*98*). Taken as a whole, the most promising avenues for advanced quantitative and predictive defect-property relationships view experimental innovations and data science not only as parallel tracks, but inextricably intertwined.

## Acknowledgments

TBB acknowledges funding from the Royal Academy of Engineering for his Research Fellowship. DSG acknowledges partial support from the National Science Foundation MRSEC Program through DMR 1720256 (IRG-1).

## References

1.  G. I. Taylor, The Mechanism of Plastic Deformation of Crystals. Part I. Theoretical. *Proc. R. Soc. A Math. Phys. Eng. Sci.* **145**, 362–387 (1934).
2.  E. Orowan, Zur Kristallplastizität. I. *Zeitschrift für Phys.* **89**, 605–613 (1934).
3.  M. Polanyi, Über eine Art Gitterstörung, die einen Kristall plastisch machen könnte. *Zeitschrift für Phys.* **89**, 660–664 (1934).
4.  J. L. Farvacque, J. C. Douehan, U. von Alpen, E. Gmelin, Screw-dislocation-induced scattering processes and acceptor states in Te. *Phys. Status Solidi.* **79**, 763–773 (1977).
5.  J. P. Hirth, R. C. Pond, Steps, dislocations and disconnections as interface defects relating to structure and phase transformations. *Acta Mater.* **44**, 4749–4763 (1996).

6.      P. B. Hirsch, R. W. Horne, M. J. Whelan, LXVIII. Direct observations of the arrangement and motion of dislocations in aluminium. *Philos. Mag.* **1**, 677–684 (1956).

7.      J. A. Venables, C. J. Harland, Electron back-scattering patterns—a new technique for obtaining crystallographic information in the scanning electron microscope. *Philos. Mag.* (1973), doi:10.1080/14786437308225827.

8.      D. J. Dingley, Diffraction from sub-micron areas using electron backscattering in a scanning electron microscope. *Scan. Electron Microsc.* **v**, 569–575 (1984).

9.      N. C. Krieger Lassen, Automatic localisation of electron backscattering pattern bands from Hough transform. *Mater. Sci. Technol.* (1996), doi:10.1179/026708396790122305.

10.      B. L. Adams, S. I. Wright, K. Kunze, Orientation imaging: The emergence of a new microscopy. *Metall. Trans. A* (1993), doi:10.1007/BF02656503.

11.      A. J. Wilkinson, G. Meaden, D. J. Dingley, High-resolution elastic strain measurement from electron backscatter diffraction patterns: New levels of sensitivity. *Ultramicroscopy* **106**, 307–313 (2006).

12.      A. J. Wilkinson, G. Meaden, D. J. Dingley, High resolution mapping of strains and rotations using electron backscatter diffraction. *Mater. Sci. Technol.* **22**, 1271–1278 (2006).

13.      T. B. Britton, A. J. Wilkinson, High resolution electron backscatter diffraction measurements of elastic strain variations in the presence of larger lattice rotations. *Ultramicroscopy*. **114**, 82–95 (2012).

14.      C. Maurice, J. H. Driver, R. Fortunier, On solving the orientation gradient dependency of high angular resolution EBSD. *Ultramicroscopy*. **113**, 171–181 (2012).

15.      D. Dingley, Progressive steps in the development of electron backscatter diffraction and orientation imaging microscopy. *J. Microsc.* **213**, 214–224 (2004).

16.      A. J. Wilkinson, T. B. Britton, Strains, planes, and EBSD in materials science. *Mater. Today*. **15**, 366–376 (2012).

17.      M. D. Vaudin, G. Stan, Y. B. Gerbig, R. F. Cook, High resolution surface morphology measurements using EBSD cross-correlation techniques and AFM. *Ultramicroscopy* (2011), doi:10.1016/j.ultramic.2011.01.039.

18.      M. D. Vaudin, Y. B. Gerbig, S. J. Stranick, R. F. Cook, Comparison of nanoscale measurements of strain and stress using electron back scattered diffraction and confocal Raman microscopy. *Appl. Phys. Lett.* (2008), doi:10.1063/1.3026542.

19.      A. J. Wilkinson, D. Randman, Determination of elastic strain fields and geometrically necessary dislocation distributions near nanoindents using electron back scatter diffraction. *Philos. Mag.* **90**, 1159–1177 (2010).

20.      J. F. Nye, Some geometrical relations in dislocated crystals. *Acta Metall.* (1953), doi:10.1016/0001-6160(53)90054-6.

21.      S. Das, F. Hofmann, E. Tarleton, Consistent determination of geometrically necessary dislocation density from simulations and experiments. *Int. J.*

*Plast.* (2018), doi:10.1016/j.ijplas.2018.05.001.

22. T. B. Britton, H. Liang, F. P. E. Dunne, A. J. Wilkinson, The effect of crystal orientation on the indentation response of commercially pure titanium: experiments and simulations. *Proc. R. Soc. a-Mathematical Phys. Eng. Sci.* **466**, 695–719 (2010).

23. T. B. Britton, J. L. R. Hickey, in *IOP Conference Series: Materials Science and Engineering* (2018).

24. A. J. Wilkinson *et al.*, Measurement of probability distributions for internal stresses in dislocated crystals. *Appl. Phys. Lett.* **105** (2014), doi:Artn 18190710.1063/1.4901219.

25. S. Kalácska, I. Groma, A. Borbély, P. D. Ispánovity, Comparison of the dislocation density obtained by HR-EBSD and X-ray profile analysis. *Appl. Phys. Lett.* (2017), doi:10.1063/1.4977569.

26. J. Jiang, T. B. Britton, A. J. Wilkinson, Measurement of geometrically necessary dislocation density with high resolution electron backscatter diffraction: Effects of detector binning and step size. *Ultramicroscopy.* **125**, 1–9 (2013).

27. J. W. Kysar, Y. Saito, M. S. Oztop, D. Lee, W. T. Huh, Experimental lower bounds on geometrically necessary dislocation density. *Int. J. Plast.* (2010), doi:10.1016/j.ijplas.2010.03.009.

28. J. Jiang, T. B. Britton, A. J. Wilkinson, Accumulation of geometrically necessary dislocations near grain boundaries in deformed copper. *Philos. Mag. Lett.* **92**, 580–588 (2012).

29. J. Jiang, T. B. Britton, A. J. Wilkinson, Evolution of dislocation density distributions in copper during tensile deformation. *Acta Mater.* **61**, 7227–7239 (2013).

30. T. J. Ruggles, D. T. Fullwood, Estimations of bulk geometrically necessary dislocation density using high resolution EBSD. *Ultramicroscopy* (2013), doi:10.1016/j.ultramic.2013.04.011.

31. A. Vilalta-Clemente *et al.*, Cross-correlation based high resolution electron backscatter diffraction and electron channelling contrast imaging for strain mapping and dislocation distributions in InAlN thin films. *Acta Mater.* **125**, 125–135 (2017).

32. N. Schäfer *et al.*, Microstrain distribution mapping on CuInSe2thin films by means of electron backscatter diffraction, X-ray diffraction, and Raman microspectroscopy. *Ultramicroscopy* (2016), doi:10.1016/j.ultramic.2016.07.001.

33. F. Javaid, E. Bruder, K. Durst, Indentation size effect and dislocation structure evolution in (001) oriented SrTiO3 Berkovich indentations: HR-EBSD and etch-pit analysis. *Acta Mater.* (2017), doi:10.1016/j.actamat.2017.07.055.

34. D. Wallis, L. N. Hansen, T. Ben Britton, A. J. Wilkinson, Dislocation Interactions in Olivine Revealed by HR-EBSD. *J. Geophys. Res. Solid Earth* (2017), doi:10.1002/2017JB014513.

35. D. Wallis, L. N. Hansen, T. Ben Britton, A. J. Wilkinson, Geometrically necessary dislocation densities in olivine obtained using high-angular resolution electron backscatter diffraction. *Ultramicroscopy* (2016), doi:10.1016/j.ultramic.2016.06.002.

36. D. Wallis, A. J. Parsons, L. N. Hansen, Quantifying geometrically necessary dislocations in quartz using HR-EBSD: Application to chessboard subgrain boundaries. *J. Struct. Geol.* (2017), , doi:10.1016/j.jsg.2017.12.012.

37. T. Zhang, J. Jiang, B. Britton, B. Shollock, F. Dunne, Crack nucleation using combined crystal plasticity modelling, high-resolution digital image correlation and high-resolution electron backscatter diffraction in a superalloy containing non-metallic inclusions under fatigue. *Proc. R. Soc. A Math. Phys. Eng. Sci.* (2016), doi:10.1098/rspa.2015.0792.

38. T. B. Britton *et al.*, Assessing the precision of strain measurements using electron backscatter diffraction - Part 2: Experimental demonstration. *Ultramicroscopy*. **135**, 126–135 (2013).

39. A. C. Milazzo *et al.*, Characterization of a direct detection device imaging camera for transmission electron microscopy. *Ultramicroscopy*. **110**, 741–744 (2010).

40. S. Vespucci *et al.*, Digital direct electron imaging of energy-filtered electron backscatter diffraction patterns. *Phys. Rev. B*. **92** (2015), doi:ARTN 20530110.1103/PhysRevB.92.205301.

41. K. Mingard, M. Stewart, M. G. Gee, S. Vespucci, C. Trager-Cowan, Practical application of direct electron detectors to EBSD mapping in 2D and 3D. *Ultramicroscopy*. **184** (2017).

42. A. Winkelmann, C. Trager-Cowan, F. Sweeney, A. P. Day, P. Parbrook, Many-beam dynamical simulation of electron backscatter diffraction patterns. *Ultramicroscopy*. **107**, 414–421 (2007).

43. A. J. Wilkinson, D. J. Dingley, The distribution of plastic deformation in a metal matrix composite caused by straining transverse to the fibre direction. *Acta Metall. Mater.* (1992), doi:10.1016/0956-7151(92)90049-K.

44. T. Ungár, Microstructural parameters from X-ray diffraction peak broadening. *Scr. Mater.* (2004), doi:10.1016/j.scriptamat.2004.05.007.

45. F. Ram, S. Zaefferer, D. Raabe, in *Journal of Applied Crystallography* (2014).

46. R. Hielscher, F. Bartel, T. B. Britton, Gazing at crystal balls - Electron backscatter diffraction indexing and cross correlation on a sphere. *eprint arXiv:1810.03211* (2018), p. arXiv:1810.03211, (available at https://ui.adsabs.harvard.edu/%5C#abs/2018arXiv181003211H).

47. R. R. Keller, R. H. Geiss, Transmission EBSD from 10 nm domains in a scanning electron microscope. *J. Microsc.* **245**, 245–251 (2012).

48. P. W. Trimby, Orientation mapping of nanostructured materials using transmission Kikuchi diffraction in the scanning electron microscope. *Ultramicroscopy*. **120**, 16–24 (2012).

49. X. Zhu *et al.*, In-depth evolution of chemical states and sub-10-nm-resolution crystal orientation mapping of nanograins in Ti (5 nm)/Au (20 nm)/Cr (3 nm) tri-layer thin films. *Appl. Surf. Sci.* (2018), doi:10.1016/j.apsusc.2018.05.042.

50.    A. B. Fanta *et al.*, Elevated temperature transmission Kikuchi diffraction in the SEM. *Mater. Charact.* (2018), doi:10.1016/j.matchar.2018.03.026.

51.    Q. Yu, M. Legros, A. M. Minor, In situ TEM nanomechanics. **40** (2015), doi:10.1557/mrs.2014.306.

52.    J. Ast, G. Mohanty, Y. Guo, J. Michler, X. Maeder, In situ micromechanical testing of tungsten micro-cantilevers using HR-EBSD for the assessment of deformation evolution. *Mater. Des.* (2017), doi:10.1016/j.matdes.2016.12.052.

53.    X. Maeder, W. M. Mook, C. Niederberger, J. Michler, Quantitative stress/strain mapping during micropillar compression. *Philos. Mag.* (2011), doi:10.1080/14786435.2010.505178.

54.    Y. Guo, J. Schwiedrzik, J. Michler, X. Maeder, On the nucleation and growth of {112̄2} twin in commercial purity titanium: In situ investigation of the local stress field and dislocation density distribution. *Acta Mater.* (2016), doi:10.1016/j.actamat.2016.08.073.

55.    T.-S. Jun *et al.*, The role of β-titanium ligaments in the deformation of dual phase titanium alloys. *Mater. Sci. Eng. A.* **746**, 394–405 (2019).

56.    D. G. Coates, Kikuchi-like reflection patterns obtained with the scanning electron microscope. *Philos. Mag.* (1967), doi:10.1080/14786436708229968.

57.    D. C. Joy, D. E. Newbury, D. L. Davidson, Electron channeling patterns in the scanning electron microscope. *J. Appl. Phys.* (1982), doi:10.1063/1.331668.

58.    R. J. Kamaladasa, Y. N. Picard, Basic Principles and Application of Electron Channeling in a Scanning Electron Microscope for Dislocation Analysis. *Microsc. Sci. Technol. Appl. Educ.* (2010).

59.    M. A. Crimp, B. A. Simkin, B. C. Ng, Demonstration of the g·b × u = 0 edge dislocation invisibility criterion for electron channelling contrast imaging. *Philos. Mag. Lett.* (2001), doi:10.1080/09500830110088755.

60.    M. E. Twigg, Y. N. Picard, Simulation and analysis of electron channeling contrast images of threading screw dislocations in 4H-SiC. *J. Appl. Phys.* (2009), doi:10.1063/1.3110086.

61.    G. Naresh-Kumar *et al.*, Imaging and identifying defects in nitride semiconductor thin films using a scanning electron microscope. *Phys. status solidi* (2012), doi:10.1002/pssa.201100416.

62.    H. Kriaa, A. Guitton, N. Maloufi, Fundamental and experimental aspects of diffraction for characterizing dislocations by electron channeling contrast imaging in scanning electron microscope. *Sci. Rep.* (2017), doi:10.1038/s41598-017-09756-3.

63.    A. J. Wilkinson, P. B. Hirsch, Electron diffraction based techniques in scanning electron microscopy of bulk materials. *Micron.* **28**, 279–308 (1997).

64.    S. Zaefferer, N. N. Elhami, Theory and application of electron channelling contrast imaging under controlled diffraction conditions. *Acta Mater.* (2014), doi:10.1016/j.actamat.2014.04.018.

65.    I. Gutierrez-Urrutia, S. Zaefferer, D. Raabe, Electron channeling contrast

imaging of twins and dislocations in twinning-induced plasticity steels under controlled diffraction conditions in a scanning electron microscope. *Scr. Mater.* (2009), doi:10.1016/j.scriptamat.2009.06.018.

66.    J. Guyon *et al.*, Sub-micron resolution selected area electron channeling patterns. *Ultramicroscopy* (2015), doi:10.1016/j.ultramic.2014.11.004.

67.    C. Lafond, T. Douillard, S. Cazottes, P. Steyer, C. Langlois, Electron CHanneling ORientation Determination (eCHORD): An original approach to crystalline orientation mapping. *Ultramicroscopy* (2018), doi:10.1016/j.ultramic.2017.12.019.

68.    H. K. Birnbaum, P. Sofronis, Hydrogen-enhanced localized plasticity-a mechanism for hydrogen-related fracture. *Mater. Sci. Eng. A* (1994), doi:10.1016/0921-5093(94)90975-X.

69.    P. B. Hirsch, R. W. Horne, M. J. Whelan, LXVIII. Direct observations of the arrangement and motion of dislocations in aluminium. *Philos. Mag.* **1**, 677–684 (1956).

70.    D. J. H. Cockayne, Weak-Beam Electron Microscopy. *Annu. Rev. Mater. Sci.* **11**, 75–95 (1981).

71.    C. J. Humphreys, Fundamental concepts of stem imaging. *Ultramicroscopy* (1981), doi:10.1016/0304-3991(81)90017-6.

72.    P. J. Phillips, M. C. Brandes, M. J. Mills, M. de Graef, Diffraction contrast STEM of dislocations: Imaging and simulations. *Ultramicroscopy*. **111**, 1483–1487 (2011).

73.    P. J. Phillips, M. J. Mills, M. De Graef, Systematic row and zone axis STEM defect image simulations. *Philos. Mag.* **91**, 2081–2101 (2011).

74.    P. J. Phillips *et al.*, Atomic-resolution defect contrast in low angle annular dark-field STEM. *Ultramicroscopy* (2012), doi:10.1016/j.ultramic.2012.03.013.

75.    P. S. Phani *et al.*, Scanning transmission electron microscope observations of defects in as-grown and pre-strained Mo alloy fibers. *Acta Mater.* **59**, 2172–2179 (2011).

76.    E. Oho *et al.*, The conversion of a field-emission scanning electron microscope to a high-resolution, high-performance scanning transmission electron microscope, while maintaining original functions. *J. Electron Microsc. Tech.* **6**, 15–30 (1987).

77.    J. Holm, R. R. Keller, Acceptance Angle Control for Improved Transmission Imaging in an SEM. *Micros. Today*. **25**, 12–19 (2017).

78.    P. G. Callahan *et al.*, Transmission scanning electron microscopy: Defect observations and image simulations. *Ultramicroscopy*. **186**, 49–61 (2018).

79.    T. Klein, E. Buhr, C. Georg Frase, TSEM: A review of scanning electron microscopy in transmission mode and its applications. *Adv. Imaging Electron Phys.* (2012), doi:10.1016/B978-0-12-394297-5.00006-4.

80.    T. Ichinokawa, S. Takashima, H. Hashimoto, S. Kimoto, Dislocation images in the high resolution scanning electron microscope AU‐Stern, R. M. *Philos. Mag. A J. Theor. Exp. Appl. Phys.* **26**, 1495–1499 (1972).

81.    B. J. Crawford, C. R. W. Liley, A simple transmission stage using the standard collection system in the scanning electron microscope. *J. Phys. E.*

(1970), doi:10.1088/0022-3735/3/6/314.

82.    W. R. McKinney, P. V. C. Hough, (United States, 1976; https://www.osti.gov/servlets/purl/7366956).

83.    J. Holm, R. R. Keller, Angularly-selective transmission imaging in a scanning electron microscope. *Ultramicroscopy* (2016), doi:10.1016/j.ultramic.2016.05.001.

84.    B. W. Caplins, J. D. Holm, R. R. Keller, Transmission imaging with a programmable detector in a scanning electron microscope. *Ultramicroscopy*. **196**, 40–48 (2019).

85.    M. Nakagawa *et al.*, Low voltage FE-STEM for characterization of state-of-the-art silicon SRAM. *J. Electron Microsc. (Tokyo).* (2002), doi:10.1093/jmicro/51.1.53.

86.    F. Grillon, Low Voltage Contrast with an SEM Transmission Electron Detector. *Microchim. Acta*. **155**, 157–161 (2006).

87.    M. R. Lee, C. L. Smith, Scanning transmission electron microscopy using a SEM: Applications to mineralogy and petrology. *Mineral. Mag.* (2006), doi:10.1180/0026461067050351.

88.    C. Sun, E. Müller, M. Meffert, D. Gerthsen, On the Progress of Scanning Transmission Electron Microscopy (STEM) Imaging in a Scanning Electron Microscope. *Microsc. Microanal.* **24**, 99–106 (2018).

89.    J. Kwon *et al.*, Characterization of dislocation structures and deformation mechanisms in as-grown and deformed directionally solidified NiAl-Mo composites. *Acta Mater.* (2015), doi:10.1016/j.actamat.2015.01.059.

90.    F. Bachmann, R. Hielscher, H. Schaeben, Grain detection from 2d and 3d EBSD data-Specification of the MTEX algorithm. *Ultramicroscopy*. **111**, 1720–1733 (2011).

91.    I. M. Robertson *et al.*, Towards an integrated materials characterization toolbox. *J. Mater. Res.* **26**, 1341–1383 (2011).

92.    T. B. Britton, V. S. Tong, J. Hickey, A. Foden, A. J. Wilkinson, AstroEBSD: exploring new space in pattern indexing with methods launched from an astronomical approach. *J. Appl. Crystallogr.* (2018), , doi:10.1107/S1600576718010373.

93.    T. B. Britton *et al.*, Tutorial: Crystal orientations and EBSD - Or which way is up? *Mater. Charact.* **117**, 113–126 (2016).

94.    S. Singh, F. Ram, M. de Graef, EMsoft: Open Source Software for Electron Diffraction/Image Simulaitons. *Microsc. Microanal.* **23:S1**, 213–232 (2017).

95.    G. Naresh-Kumar *et al.*, Coincident electron channeling and cathodoluminescence studies of threading dislocations in GaN. *Microsc. Microanal.* (2014), doi:10.1017/S1431927613013755.

96.    A. R. Rao, Future directions in industrial machine vision: A case study of semiconductor manufacturing applications. *Image Vis. Comput.* (1996), doi:10.1016/0262-8856(95)01035-1.

97.    A. J. Wilkinson, D. M. Collins, Y. Zayachuk, R. Korla, A. Vilalta-Clemente, Applications of multivariate statistical methods and simulation libraries to analysis of electron backscatter diffraction and transmission

          Kikuchi diffraction datasets. *Ultramicroscopy* (2019), doi:10.1016/j.ultramic.2018.09.011.

98. F. de la Peña, U. of Lille, ; Vidar Tonaas Fauske; Pierre Burdet; Eric Prestat; Petras Jokubauskas; Magnus Nord; Tomas Ostasevicius; Katherine E. MacArthur; Mike Sarahan; Duncan N. Johnstone; Joshua Taillon; Alberto Eljarrat; Vadim Migunov; Jan Caron; Tom Furnival; Stefano Mazzucco; Skorikov, hyperspy/hyperspy v1.4.1, , doi:10.5281/zenodo.1469364.

99. T. B. Britton, A. J. Wilkinson, Stress fields and geometrically necessary dislocation density distributions near the head of a blocked slip band. *Acta Mater.* **60**, 5773–5782 (2012).

100. H. Yu, J. Liu, P. Karamched, A. J. Wilkinson, F. Hofmann, Mapping the full lattice strain tensor of a single dislocation by high angular resolution transmission Kikuchi diffraction (HR-TKD). *Scr. Mater.* **164**, 36–41 (2019).

101. P. J. Phillips, M. J. Mills, M. De Graef, Systematic row and zone axis STEM defect image simulations. *Philos. Mag.* (2011), doi:10.1080/14786435.2010.547526.

102. P. G. Callahan, B. B. Haidet, D. Jung, G. G. E. Seward, K. Mukherjee, Direct observation of recombination-enhanced dislocation glide in heteroepitaxial GaAs on silicon. *Phys. Rev. Mater.* **2**, 81601 (2018).

103. J. C. Stinville *et al.*, Dislocation Dynamics in a Nickel-Based Superalloy via In-Situ Transmission Scanning Electron Microscopy. *Acta Mater.* (2019), doi:10.1016/J.ACTAMAT.2018.12.061.

104. S. K. Makineni *et al.*, Correlative Microscopy—Novel Methods and Their Applications to Explore 3D Chemistry and Structure of Nanoscale Lattice Defects: A Case Study in Superalloys. *JOM* (2018), doi:10.1007/s11837-018-2802-7.

**Author biographies**

**Daniel Gianola** is an Associate Professor of Materials at the University of California, Santa Barbara. He leads a research group specializing in deformation at the micro- and nanoscale, particularly using *in situ* nanomechanical testing techniques. His group also develops novel combinatorial synthesis methods and high-throughput characterization approaches to accelerate alloy design. Prof. Gianola is currently the faculty director of the Microscopy and Microanalysis Facility at UCSB, which is a state-of-the-art shared experimental facility on campus. Prof. Gianola is the recipient of the National Science Foundation CAREER, Department of Energy Early Career, and TMS Early Career Faculty Fellow awards.

**Dr. Ben Britton** is a senior lecturer (~associate professor) at Imperial College London. He leads the experimental micromechanics group, which develops new understanding of the deformation behaviour of materials used in aerospace, oil & gas, and nuclear power applications. To aid in characterisation of deformation processes, the group have been developing advances in EBSD and ECCI based mapping including high angular resolution EBSD and open source tools like AstroEBSD and combining these with in situ deformation.

**Stefan Zaefferer** studied physical metallurgy and metal physics at the TU Clausthal in Germany where he also obtained his PhD degree. During PhD and post doc times in Paris and in Kyoto he developed the computer program TOCA for on-line indexing of TEM and SEM diffraction patterns and microscope control. Since 2000 he is head of the research group „Microscopy and Diffraction" at the Max-Planck-Institut für Eisenforschung (MPIE) and university lecturer at the RWTH Aachen as well as guest professor at various universities. His research interests span the development of tools for electron microscopy and investigation of microstructure mechanisms in various structural and functional materials.

**Figures**

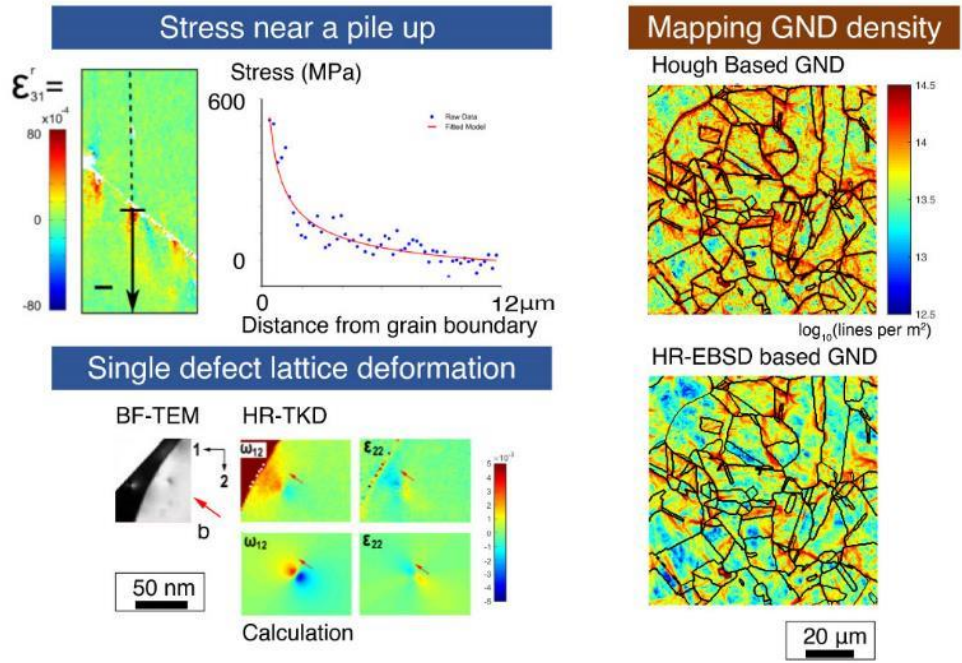

**Figure 1.** EBSD measurements of defects and defect properties, including: High Angular-EBSD stress near a dislocation pile up and a demonstration of the Eshelby Frank and Nabarro prediction that the stress decays as $distance^{-1/2}$. Reprinted from (*99*) with permission from Elsevier; a demonstration that HR-EBSD provides increased resolution to map the accumulation of dislocations with respect to microstructural features (shown here in copper) (*23*); and High Angular Resolution TKD showing mapping of the stress and rotation fields around a single dislocation (*100*).

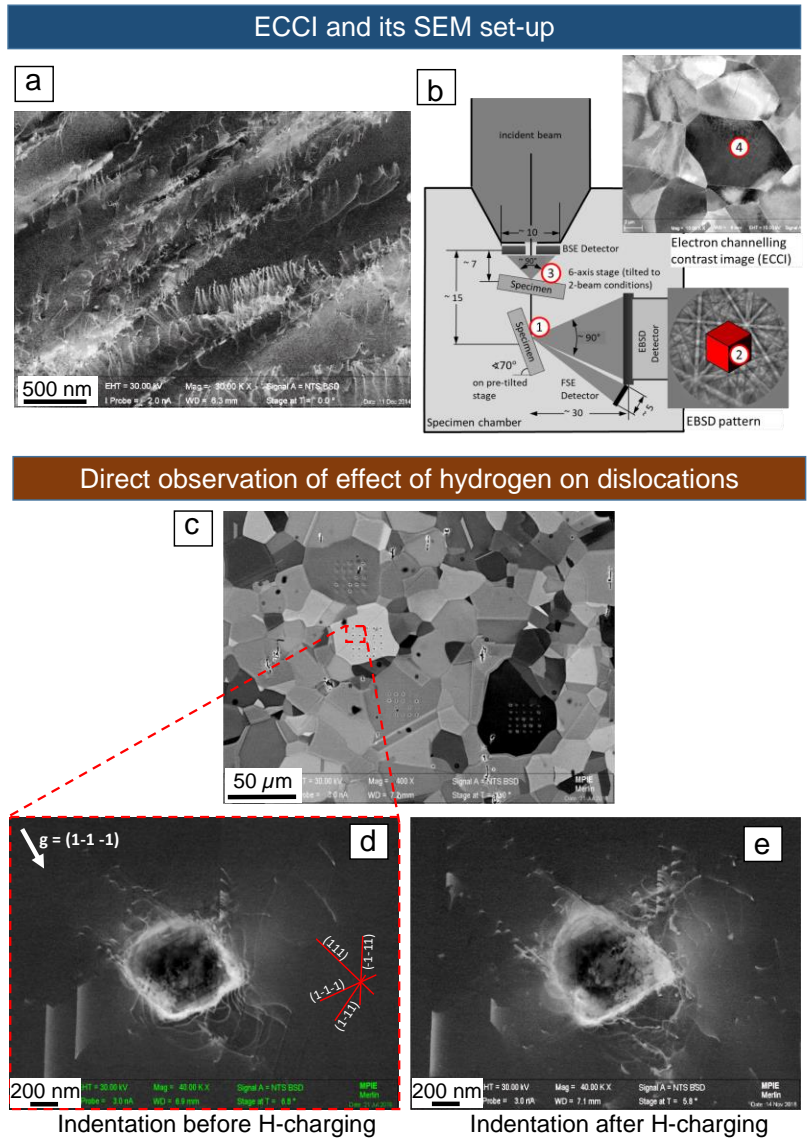

**Figure 2.** ECC imaging of crystalline defects. (a) ECC image of a lightly deformed high entropy alloy, showing planar bands of dislocations. (b) EBSD-assisted ECCI to determine crystallographic orientation and corresponding diffraction conditions. (c-e) Hydrogen-dislocation interactions in a high-Mn austenitic steel sample, with ECCI imaging of dislocations in the vicinity of an indentation (d) prior to and (e) following H-charging.

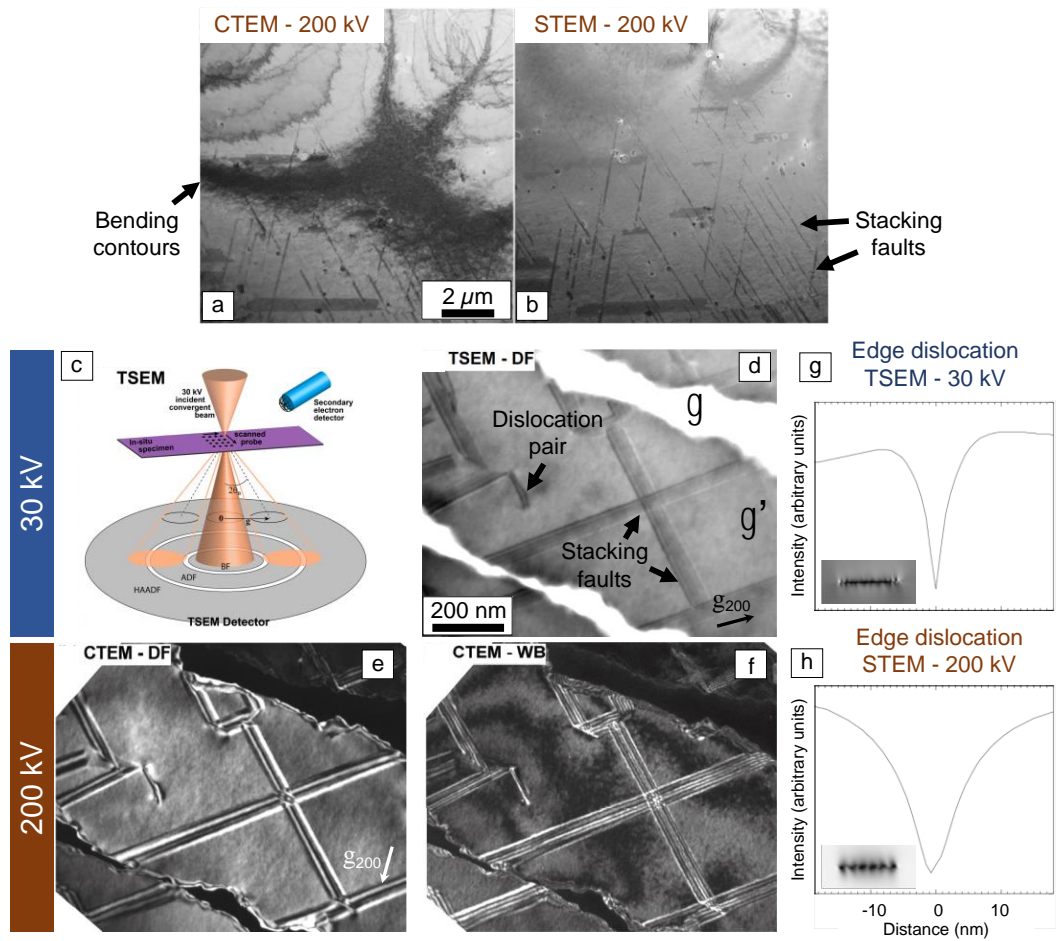

**Figure 3.** Imaging of crystalline defects using a variety of diffraction-contrast electron microscopy. (a,b) A comparison of faulted Ni-based superalloy in a severely bent foil imaged using (a) CTEM and (b) STEM at 200 kV, demonstrating the muting of bending contours via STEM imaging. Reprinted by permission of the publisher (Taylor & Francis Ltd, http://www.tandfonline.com) (*101*). (c) Schematic of TSEM (or STEM-in-SEM) imaging conditions at 30 kV using a transmission detector in an SEM. (d-f) Comparison of a Co-based superalloy following high temperature creep deformation and possessing stacking faults and dislocations, as imaged using (d) 30 kV TSEM, (e) 200 kV CTEM, and (f) 200 kV weak-beam (WB) CTEM. Note the similarity in fringe density within the stacking faults for TSEM and WB images. Reprinted from (*78*) with permission from Elsevier. (g,h) Dynamical diffraction simulations showing integrated intensities across edge dislocations in Co for (g) 30 kV and (h) 200 kV STEM probes, demonstrating the localization of dislocation contrast at low primary beam energies owing to the smaller extinction distances. Simulations courtesy of P. Callahan.

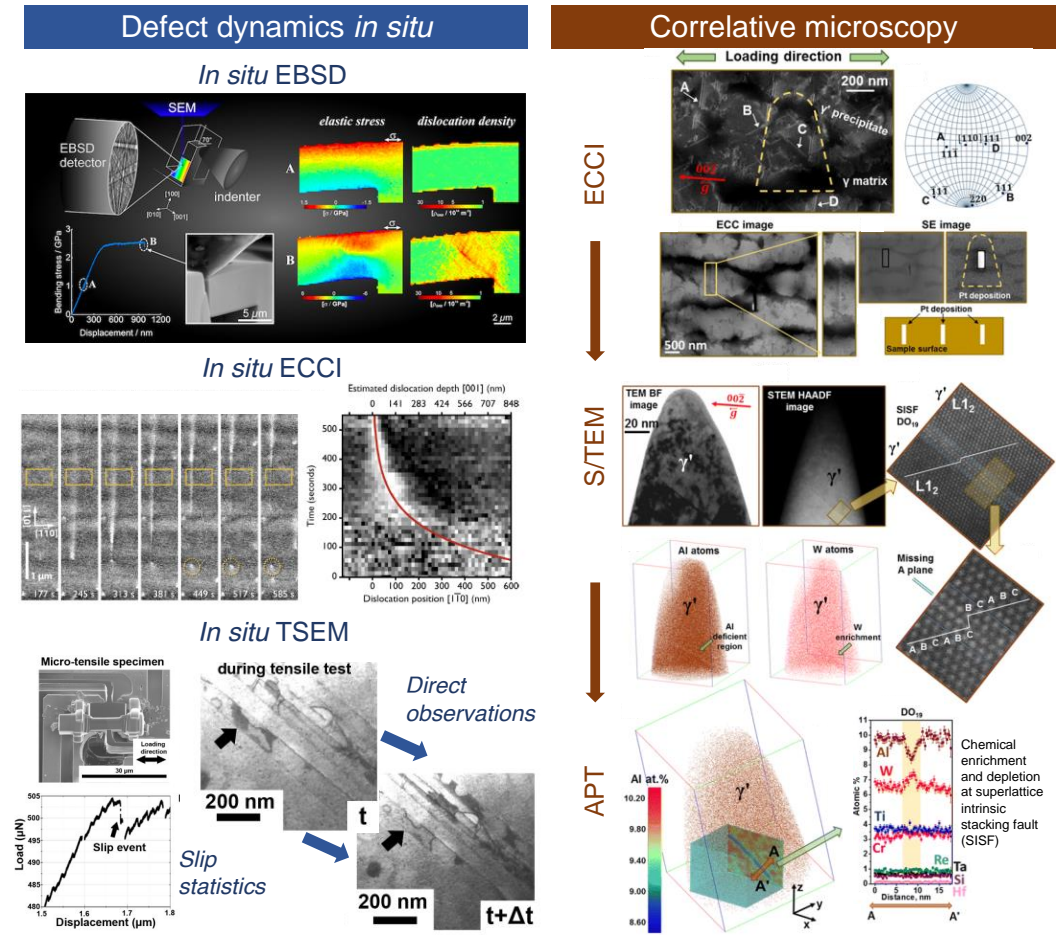

**Figure 4.** Applications of modern defect imaging and diffraction modes to *in situ* observations and correlative microscopy. Examples of *in situ* results (left panel) are shown for EBSD (reprinted from (*52*) with permission from Elsevier), ECCI (reprinted figure with permission from (*102*). Copyright (2019) by the American Physical Society), and TSEM (reprinted from (*103*) with permission from Elsevier). A highlight of correlative microscopy (right panel) demonstrates the use of ECCI to identify defects for site-specific investigations, including S/TEM and atom probe tomography (APT) methods to characterize structural and chemical details, respectively, at a superlattice intrinsic stacking fault in a CoNi-based superalloy (*104*).